\documentclass{svjour3}
\smartqed  \usepackage{graphicx}
\usepackage{fix-cm}
\usepackage{amsmath}

\journalname{General Relativity and Gravitation}

\begin{document}

\title{Radiating stars with generalised Vaidya atmospheres}

\author{S. D. Maharaj \and G. Govender  \and  M. Govender}

\institute{
S. D. Maharaj \and G. Govender       \and  M. Govender \at
Astrophysics and Cosmology Research Unit,
 School of Mathematical Sciences,
 University of KwaZulu-Natal, Private Bag X54001,Durban 4000, South Africa
\\
\email{maharaj@ukzn.ac.za} \\ \\
}
\date{Received: date / Accepted: date}

\maketitle

\begin{abstract}
We model the gravitational behaviour of a radiating star when the
exterior geometry is the generalised Vaidya spacetime. The interior
matter distribution is shear-free and undergoing radial heat flow.
The exterior energy momentum tensor is a superposition of a null
fluid and a string fluid. An analysis of the junction conditions at
the stellar surface shows that the pressure at the boundary depends
on the interior heat flux and the exterior string density. The
results for a relativistic radiating star undergoing nonadiabatic
collapse are obtained as a special case. For a particular model we
demonstrate that the radiating fluid sphere collapses without the
appearance of the horizon at the boundary.


\end{abstract}

\section{Introduction}
The study of radiating stars in the context of general relativity
has generated much interest in researchers because of the variety of
applications in relativistic astrophysics. These studies are
important as they enable us to investigate physical features such as
surface luminosity, dynamical stability, particle production at the
stellar surface, relaxation effects, causal temperature gradients
and other thermodynamical processes. Some relevant references
investigating these issues are given by Di Prisco et al.
\cite{b1}, Govender et al. \cite{b2}, Herrera et al.
\cite{b3} and Pinheiro and Chan \cite{b4}. Relativistic radiating
stars are also important in the process of gravitational collapse,
describing the final state of stars, formation of singularities and
black hole physics, in four and higher dimensions. Recent
investigations in this regard are contained in the works of Goswami
and Joshi \cite{b5}, Joshi \cite{b6} and Madhav et al.
\cite{b7}. In particular, the validity of the cosmic censorship
conjecture can be tested in this physical scenario.

The model of a relativistic radiating star undergoing dissipation
was completed by Santos \cite{b8} by analysing the junction
conditions at the stellar surface. By matching a shear-free interior
spacetime to the radiating Vaidya exterior spacetime, he showed that
at the surface the pressure is nonvanishing and proportional to the
heat flux. Subsequently several explicit relativistic radiating
stellar models have been found by investigating the appropriate
boundary condition. Kramer \cite{b9} and Maharaj and Govender
\cite{b10} generated nonstatic radiating spheres from a static model
by allowing certain parameters to become functions of time. Kolassis
et al. \cite{b11} and Thirukkanesh and Maharaj \cite{b12}
assumed geodesic fluid trajectories to produce new radiating models.
In the approach of De Oliviera et al. \cite{b13} and Nogueira
and Chan \cite{b14} the model has an initial static configuration
before the radiating sphere starts gradually to collapse. Exact
solutions for shear-free interiors which are conformally flat
generate radiating stellar models as shown by Herrera et al.
\cite{b15}, Herrera et al. \cite{b16}, Maharaj and Govender
\cite{b17} and Misthry et al. \cite{b18}. Stellar models
which are radiating with nonzero shear are difficult to analyse
because of the complexity of the boundary condition. However even in
this case there have been advances in obtaining exact solutions.
Particular exact models have been found by Naidu et al.
\cite{b19}, Rajah and Maharaj \cite{b20} and Pinheiro and Chan
\cite{b21}.

In this paper we seek to generalise the Santos junction conditions
by matching a shear-free interior spacetime to the generalised
Vaidya exterior spacetime. The energy momentum tensor of the
generalised Vaidya spacetime may be interpreted as a superposition
of two fluids, a null dust and a null string fluid. The physical
properties of the generalised Vaidya spacetime have been discussed
by Husain \cite{b22} and Wang and Wu \cite{b23}. Glass and Krisch
\cite{b24} have interpreted the exterior spacetime as a
superposition of two fluids outside a relativistic star, the
original Vaidya null fluid and a new null fluid composed of strings.
By assuming diffusive transport for the string fluid Glass and
Krisch \cite{b25} found new solutions to Einstein's equations with
transverse stresses. Physically reasonable energy transport
mechanisms have been generated by Krisch and Glass \cite{b26} in the
stellar interior with the generalised Vaidya metric as the exterior
spacetime. These investigations, and other treatments, have largely
focussed on physical processes in the exterior of the stellar model
with a generalised Vaidya atmosphere. To fully describe a radiating
stellar model requires generation of the junction conditions at the
stellar surface.

We follow the convention that the coupling constant $\frac{8\pi
G}{c^4}$ and the speed of light $c$ are unity; the metric has
signature $(-\,+\,+\,+)$. In Sect. 2 we present the field equations for
the interior and exterior spacetimes. In Sect. 3 the matching of the
interior and exterior spacetimes across the stellar surface is
outlined. The new set of junction conditions are derived for the
generalised Vaidya spacetime. We indicate how the new junction
conditions generalise the junction conditions previously derived by
Santos \cite{b8}. The physical significance of our new result is
highlighted in terms of a string fluid. We also consider the new
junction condition in the context of conservation of momentum flux
across the stellar boundary. In Sect. 4 we generate a particular model
and demonstrate that the new generalised junction condition has a
solution. In this model it is possible for the radiating fluid
sphere to collapse without the appearance of a boundary. The results of
this paper are briefly summarised in Sect. 5.

\section{Field Equations}
Spacetime needs to be divided into two distinct regions, the
interior spacetime $\cal M^{-}$ and the exterior spacetime $\cal
M^{+}$ for a stellar model. The boundary of the star $\Sigma$ serves
as the matching surface for $\cal M^{-}$ and $\cal M^{+}$. The
boundary or stellar surface is a timelike three-dimensional
hypersurface. We require for the first junction condition that
\begin{equation}
(ds_{+}^2)_{\Sigma}=(ds_{-}^2)_{\Sigma}=ds^{2}_{\Sigma}\label{metricrelation}
\end{equation}
so that the line elements match on the boundary $\Sigma$. The second
junction condition is generated by the continuity of the extrinsic
curvature of $\Sigma$ across the boundary given by
\begin{equation}
(K_{\alpha \beta}^+)_{\Sigma} = (K_{\alpha
\beta}^{-})_{\Sigma}\label{junc2}
\end{equation}
Note that the junction conditions $(\ref{metricrelation})$ and
$(\ref{junc2})$ are equivalent to the Lichnerowicz \cite{b27} and O'
Brien and Synge \cite{b28} junction conditions.

The line element for the interior manifold $\cal M^-$ is given by
\begin{equation}
ds^2 = -A^2(t, r)dt^2 + B^2(t, r)[dr^2 + r^2(d\theta^2 +
\sin^2\theta d\phi^2)] \label{abmet}
\end{equation}
 in comoving and isotropic coordinates. The interior spacetime is
expanding and accelerating but is shear-free. A physically relevant
interior matter distribution that is consistent with $(\ref{abmet})$
is given by
\begin{equation}
T_{ab}^-=(\mu+p)u_{a}u_{b}+pg_{ab}+q_{a}u_{b}+q_{b}u_{a}\label{matten}
\end{equation}
 where $\mu$ is the energy density, $p$ is the isotropic pressure,
$q_a$ is the radial heat flux vector and
$u^a=\frac{1}{A}\delta_{0}^{a}$ is the comoving fluid four-velocity.
The Einstein field equations $G_{ab}^-=T_{ab}^-$ for the interior
manifold $\cal M^-$ are given by
\begin{subequations}
\label{0sheareinfield}
\begin{eqnarray}
\mu &=& 3\frac{\dot{B}^2}{A^2B^2}-
\frac{1}{B^2}\left(2\frac{B^{\prime\prime}}{B}
-\frac{B^{\prime2}}{B^2}
+\frac{4}{r}\frac{B^{\prime}}{B}\right)\label{0sheareinfield1}\\
p &=& \frac{1}{A^2}\left(-2\frac{\ddot{B}}{B} -\frac{\dot{B}^2}{B^2}
+ 2\frac{\dot{A}}{A} \frac{\dot{B}}{B}\right) \nonumber\\
& & + \frac{1}{B^2}\left(\frac{B^{\prime 2}}{B^2} +
2\frac{A^{\prime}}{A}\frac{B^{\prime}}{B} +
\frac{2}{r}\frac{A^{\prime}}{A} +
\frac{2}{r}\frac{B^{\prime}}{B}\right)\label{0sheareinfield2}\\
p &=& -2\frac{\ddot{B}}{A^{2}B} +
2\frac{\dot{A}}{A^3}\frac{\dot{B}}{B} -
\frac{\dot{B}^2}{A^2B^2} + \frac{1}{r}\frac{A^{\prime}}{AB^2}\nonumber\\
& & + \frac{1}{r}\frac{B^{\prime}}{B^3} +
\frac{A^{\prime\prime}}{AB^2} - \frac{B^{\prime 2}}{B^4} +
\frac{B^{\prime\prime}}{B^3}\label{0sheareinfield3}\\
q &=& -\frac{2}{AB^2}\left(-\frac{\dot{B}^{\prime}}{B} +
\frac{B^{\prime}\dot{B}}{B^2} +
\frac{A^{\prime}}{A}\frac{\dot{B}}{B}\right)\label{0sheareinfield4}
\end{eqnarray}
\end{subequations}
where dots and primes denote differentiation with respect to $t$ and
$r$ respectively. The results
$(\ref{metricrelation})$-$(\ref{0sheareinfield})$ were first
obtained by Santos \cite{b8}.

The line element for the exterior manifold $\cal M^+$ is taken to be
\begin{equation}
ds^2 = - \left(1 - 2\frac{ m(v,\sf{r})}{\sf{r}}\right)dv^2 -2 dv
d{\sf{r}} + {\sf{r^2}} (d\theta^2 +\sin^2\theta
d\phi^2)\label{genvadmet}
\end{equation}
 where $m(v, \sf{r})$ is the mass function, and is related to the
gravitational energy within a given radius $\sf{r}$ (Lake and
Zannias \cite{b29}, Poisson and Israel \cite{b30}). This metric is
often called the generalised Vaidya spacetime since it reduces to
the Vaidya spacetime when $m=m(v)$ which is the mass of the star as
measured by an observer at infinity. It has been demonstrated by
Husain \cite{b22} and Wang and Wu \cite{b23} that an energy momentum
tensor consistent with $(\ref{genvadmet})$ is
\begin{eqnarray}
T_{ab}^+ &=& T_{ab}^{(n)}+T_{ab}^{(m)}\nonumber\\
&=& \varepsilon l_{a}l_{b} +
\left(\rho+P\right)\left(l_{a}n_{b}+l_{b}n_{a}\right)+Pg_{ab}\label{exmetten3}
\end{eqnarray}
which represents a superposition of a null dust and a null string
fluid. In general $T_{ab}^+$ represents a Type \textrm{II} fluid as
defined by Hawking and Ellis \cite{b31}. The null vector $l^a$ is a
double null eigenvector of the energy momentum tensor $T_{ab}^{+}$.
The weak and strong energy conditions, and the dominant energy
conditions are satisfied for proper choices of the mass function
$m(v, {\sf{r}})$. In $(\ref{exmetten3})$ we have introduced the two
null vectors
\begin{equation}
l_{a} = \delta_{a}^0,\quad \quad \quad \quad n_{a} =
\frac{1}{2}\left[1-2\frac{m(v,
\sf{r})}{\sf{r}}\right]\delta_{a}^{0}+\delta_{a}^{1}\label{exnullvecs}
\end{equation}
where $l_{a}l^{a}=n_{a}n^{a}=0$ and $l_{a}n^{a}=-1$. The Einstein
field equations $G_{ab}^+=T_{ab}^+$ for the exterior manifold $\cal
M^+$ are then given by
\begin{subequations}
\label{exeinfield}
\begin{eqnarray}
\varepsilon &=& -2\frac{m_v}{\sf{r}^2}\label{exeinfield1}\\
\rho &=& 2\frac{m_{\sf{r}}}{\sf{r}^2}\label{exeinfield2}\\
P &=& -\frac{m_{\sf{r}\sf{r}}}{\sf{r}}\label{exeinfield3}
\end{eqnarray}
\end{subequations}
where we have used the notation
\begin{equation}
m_v = \frac{\partial m}{\partial v},\quad\quad\quad
m_{{\sf{r}}}=\frac{\partial m}{\partial {\sf{r}}}\nonumber
\end{equation}
We interpret $\varepsilon$ as the density of the null dust
radiation; $\rho$ and $P$ are the null string density and null
string pressure, respectively. Note that
$(\ref{exmetten3})$-$(\ref{exeinfield3})$ were derived by Wang and
Wu \cite{b23}.

\section{Generalised Santos conditions}

It is possible to match the spacetimes $(\ref{abmet})$ and
$(\ref{genvadmet})$ across the boundary $\Sigma $. Since the
derivation is similar to the Santos \cite{b8} treatment we provide
only an outline of the argument for our more general case with
$m=m(v, \sf{r})$. The intrinsic metric to the hypersurface $\Sigma$
is defined by
\begin{equation}
ds_{\Sigma}^2 = -d \tau^{2}+{\cal{Y}}^{2} \left(d\theta^{2} +\sin
^{2}\theta d\phi^{2}\right)\label{surfacemet}
\end{equation}
For the interior spacetime $\cal M^{-}$ we obtain
\begin{subequations}
\label{surfcons}
\begin{eqnarray}
A(r_{\Sigma}, t)dt &=& d\tau \label{surfcon1}\\
r_{\Sigma}B(r_{\Sigma}, t) &=& \cal Y (\tau)\label{surfcon2}
\end{eqnarray}
\end{subequations}
For the exterior region $\cal M^+$ we generate the results
\begin{subequations}
\label{extsurfcons}
\begin{eqnarray}
{\sf{r}}_{\Sigma}(v) &=& \cal Y (\tau)\label{extsurfco1}\\
\left(1-\frac{2m}{\sf{r}}+2\frac{d\sf{r}} {dv}\right)_{\Sigma} &=&
\left(\frac{dv}{d\tau}\right)^{-2}_{\Sigma}\label{vdoteqn}
\end{eqnarray}
\end{subequations}
Equations $(\ref{surfcons})$ and $(\ref{extsurfcons})$ correspond to
the first junction condition $(\ref{metricrelation})$. Observe that
the quantity $\tau$ was defined on the surface $\Sigma$ as an
intermediate variable. On eliminating $\tau$ we have
\begin{subequations}
\label{necsufcons}
\begin{eqnarray}
A(r_{\Sigma}, t)dt &=& \left(1-\frac{2m}{\sf{r}
_{\Sigma}}+2\frac{d\sf{r} _{\Sigma}} {dv}\right)^{1/2}dv\label{necsufcon1}\\
{\sf{r}}_{\Sigma}(v) &=& rB({\sf{r}}_{\Sigma}, t)\label{necsufcon2}
\end{eqnarray}
\end{subequations}
Equations $(\ref{necsufcons})$ are the necessary and sufficient
conditions for the first junction condition $(\ref{metricrelation})$
to be valid.

The intrinsic curvature for the interior spacetime $\cal M^{-}$ has
the form
\begin{subequations}
\label{intcurv}
\begin{eqnarray}
K_{11}^{-} &=&
\left(-\frac{1}{B}\frac{A^{\prime}}{A}\right)_{\Sigma}\label{11in}\\
K_{22}^{-} &=& \left[r(rB)^{\prime}\right]_{\Sigma}\label{22in}\\
K_{33}^{-}&=& \sin^2\theta K_{22}^{-}\label{33in}
\end{eqnarray}
\end{subequations}
The extrinsic curvature for the exterior spacetime $\cal M^{+}$ has
the form
\begin{subequations}
\label{extcurvcomps}
\begin{eqnarray}
K_{11}^+ &=&
\left[\frac{\tilde{v}\hspace{-1mm}\raisebox{.5ex}{$\tilde{}$}}{\tilde{v}}-\tilde{v}\frac{m}{\sf{r}^2}+
\tilde{v}\frac{m_{{\sf{r}}}}
{\sf{r}}\right]_{\Sigma}\label{11out}\\
K_{22}^+ &=&
\left[\tilde{v}\left(1-\frac{2m}{{\sf{r}}}\right){\sf{r}}
+{\sf{r}}\tilde{{\sf{r}}}\right]_{\Sigma}\label{22out}\\
K_{33}^+ &=& \sin^2 \theta K_{22}^+\label{33out}
\end{eqnarray}
\end{subequations}
where $\tilde{{\sf{r}}} = \frac{d{\sf{r}}}{d\tau}$ and
$\tilde{v}=\frac{dv}{d\tau}$. Observe the appearance of the term
containing $m_{\sf{r}}$ in $K_{11}^{+}$ which does not exist in the
treatment of Santos \cite{b8}. As we shall see later this has a
profound effect on the physics of the model. Equations
$(\ref{intcurv})$ and $(\ref{extcurvcomps})$ correspond to the
second junction condition $(\ref{junc2})$. The mass profile in terms
of the metric functions can be generated by eliminating ${\sf{r}}$,
$\tilde{{\sf{r}}}$ and $\tilde{v}$. We observe that
\begin{equation}
m(v, {\sf{r}})= \left[\frac{rB}{2}\left(1+r^2 \frac{\dot{B}^2}{A^2}
-
\frac{1}{B^2}(B+rB^{\prime})^2\right)\right]_{\Sigma}\label{stdmass}
\end{equation}
 which is the total gravitational energy contained within the
stellar surface $\Sigma$. We also establish the relationship
\begin{equation}
\left(-\frac{1}{B}\frac{A^{\prime}}{A}\right) =
\left(1+r\frac{B^{\prime}}{B}
+r\frac{\dot{B}}{A}\right)^{-1}\nonumber\\
\nonumber\\
\end{equation}
\begin{equation}
\times\left[\frac{m_{\sf{r}}}{\sf{r}}+\frac{B^{\prime}}{B^2}+\frac{r}{2}\frac{B^{\prime
2}}{B^3}-\frac{r}{2}\frac{\dot{B}^2}{BA^2}-\frac{1}{A}\left(r\frac{\dot{B}^{\prime}}{B}
+r\frac{\ddot{B}}{A}-r\frac{B^{\prime}\dot{B}}{B^2}-r\dot{B}\frac{\dot{A}}{A^2}\right)\right]_{\Sigma}\label{relationship}
\end{equation}
This expression may be simplified further: multiply with
$1+r\frac{B^{\prime}}{B}+r\frac{\dot{B}}{A}$ and utilise
$(\ref{0sheareinfield2})$ and $(\ref{0sheareinfield4})$. We then
arrive at the result
\begin{equation}
p=\left(qB-2\frac{m_{{\sf{r}}}}{r^2B^2}\right)_{\Sigma}\label{rawresult}
\end{equation}\\
\noindent which  generalises the junction condition of Santos
\cite{b8}. Hence we have demonstrated that the junction conditions
$(\ref{junc2})$ are equivalent to
\begin{subequations}
\label{juncsetfin}
\begin{eqnarray}
m(v, {\sf{r}}) &=& \left[\frac{rB}{2}\left(1+r^2
\frac{\dot{B}^2}{A^2} -
\frac{1}{B^2}(B+rB^{\prime})^2\right)\right]_{\Sigma}\label{res1}\\
p &=&
\left(qB-2\frac{m_{{\sf{r}}}}{{\sf{r}}^2}\right)_{\Sigma}\label{res2}
\end{eqnarray}
\end{subequations}
Equations $(\ref{juncsetfin})$ are the necessary and sufficient
conditions for the second junction condition $(\ref{junc2})$ to be
valid. We point out that the mathematical approach and procedure
that we have followed is similar to Santos \cite{b8}. However in our
case the external stellar atmosphere is the generalised Vaidya
spacetime. The form of the equations
$(\ref{surfacemet})$-$(\ref{res2})$ can be related to the equations
of Santos \cite{b8} since they have a similar structure. The fact
that $m=m(v, {\sf{r}})$ fundamentally affects the final result. The
equations for the extrinsic curvature $(\ref{11out})$, the matching
condition $(\ref{relationship})$ and junction condition
$(\ref{rawresult})$ are fundamentally different.\\

We have generated the relationships $(\ref{necsufcons})$ and
$(\ref{juncsetfin})$ so that the junction conditions
$(\ref{metricrelation})$ and $(\ref{junc2})$ are satisfied for the
shear-free interior spacetime $(\ref{abmet})$ and the generalised
Vaidya exterior spacetime $(\ref{genvadmet})$ across the
hypersurface $\Sigma$. This generalises the Santos \cite{b8} result
for a relativistic radiating star when $m=m(v)$. Observe that when
$m$ depends on the coordinate $v$ only then $(\ref{res2})$ becomes
\begin{equation}
p=qB\label{santos}
\end{equation}
at the boundary $\Sigma$, which is the earlier Santos junction
condition. When $(\ref{santos})$ is valid then the pressure $p$ on
the boundary depends only on the heat flux $q$. We have shown here
that if $m=m(v, {\sf{r}})$ then $(\ref{res2})$ is valid, and the
pressure $p$ on the boundary depends on the heat flux $q$ and the
gradient $m_{{\sf{r}}}(v,{\sf{r}})$.

The generalised Vaidya spacetime has physical significance and
contains many known solutions of the Einstein field equations with
spherical symmetry. It contains the monopole solution, the de Sitter
and Anti-de Sitter solutions, the charged Vaidya solution and the
radiating dyon solution. The physical features and the energy
momentum complexes, that provide acceptable energy momentum
distributions for these systems, have been studied by Barriola and
Vilenkin \cite{b32}, Chamorro and Virbhadra \cite{b33}, Virbhadra
\cite{b34}-\cite{b36} and Yang \cite{b37}. Glass and Krisch
\cite{b24}-\cite{b25} and Krisch and Glass \cite{b26} have
interpreted the generalised Vaidya spacetime to represent a
superposition of an atmosphere composed of two fluids: a string
fluid and a null dust fluid. This atmosphere may model several
physical situations at different distance scales, eg. the exterior
regions of black holes (distance scale of multiples of the
Schwarzschild radius) and globular clusters containing a component
of dark matter (distance scale of the order of parsecs). The
additional term $2\frac{m_{{\sf{r}}}}{{\sf{r}}^2}$ in the boundary
condition $(\ref{res2})$ arises from the matching at the surface
$\Sigma$. This quantity has physical significance and can be
interpreted as a particular contribution from the energy momentum
tensor. We observe that the term $2\frac{m_{{\sf{r}}}}{{\sf{r}}^2}$
in $(\ref{res2})$ is the same quantity as that in
$(\ref{exeinfield2})$. Therefore we may interpret the quantity
$2\frac{m_{\sf{r}}}{\sf{r}^2}$ as the string density $\rho$.

We can therefore write $(\ref{res2})$ in the more transparent form
\begin{equation}
p=\left[qB-\rho\right]_{\Sigma}\label{densres}
\end{equation}
at the boundary $\Sigma$. Consequently for a radiating star with
outgoing dissipation in the form of radial heat flow, with the
generalised Vaidya spacetime as the exterior, the pressure on the
surface depends on the interior heat flux $q$ and the exterior
string density $\rho$. The appearance of the quantity $\rho$ in
$(\ref{densres})$ allows for more general behaviour that was the
case in the Santos \cite{b8} treatment. From $(\ref{santos})$ we
observe that $q=0$ implies that $p=0$ on $\Sigma$ and the exterior
manifold $\cal M^+$ must be the Schwarzschild exterior metric with
$m$ being constant. In $(\ref{densres})$ we note that we obtain the
Schwarzschild exterior geometry when $q=0=\rho$. However it is clear
from $(\ref{densres})$ that when $qB=\rho$ then $p=0$ on $\Sigma$
and the exterior spacetime remains the generalised Vaidya spacetime
with $m=m(v, {\sf{r}})$. In addition, when $q=0$ then $p=-\rho$ on
$\Sigma$ and the interior is not radiating.

It is possible to provide a physical interpretation of our result by
consideration of the momentum flux across the boundary $\Sigma$.
Since the quantity $(\ref{res1})$ represents the total gravitational
energy for a sphere of radius $r$ within $\Sigma$ we can write $m(v,
{\sf{r}})=m(t, r)$. Partially differentiating $(\ref{res1})$
eventually leads to the result
\begin{equation}
p_{\Sigma}=-\frac{2}{r^2}\tilde{v}^2m_v -
\frac{2}{r^2}m_{{\sf{r}}}\label{prelation}
\end{equation}
\noindent which reduces to the corresponding Santos \cite{b8}
equation when $m=m(v)$. The radial flux of momentum across the
hypersurface $\Sigma$ is defined by
\begin{equation}
F^{\pm}=e_{0}^{\pm a}n^{\pm b}T_{ab}^{\pm}\label{momfluxdefn}
\end{equation}
where $e_{0}^{\pm a}$ and $n^{\pm b}$ are vectors which are
respectively tangent and normal to $\Sigma$. For conservation of
momentum flux across $\Sigma$ we must have
\begin{equation}
F^+=F^-\label{momcons}
\end{equation}
In the interior manifold $\cal M^{-}$  we can generate the quantity
\begin{equation}
F^{-} = -qB\label{mfin}
\end{equation}
In the exterior manifold $\cal M^+$ we can produce the quantity
\begin{equation}
F^+ = \frac{2}{r^2}\tilde{v}^2m_v\label{fplus}
\end{equation}
 Then equations $(\ref{prelation})$-$(\ref{fplus})$ yield the result
\begin{equation}
p_{\Sigma} =
\left(qB-\frac{2}{r^2}\frac{m_{{\sf{r}}}}{B^2}\right)_{\Sigma}\nonumber
\end{equation}
 which is the same as $(\ref{res2})$. Therefore the junction
condition $(\ref{res2})$ corresponds to the conservation of the
radial flux of momentum across the hypersurface $\Sigma$. It
represents the local conservation of momentum.

\section{A particular model}
To illustrate the utility of the generalised Santos condition we
consider a specific example relating to horizons. For simplicity we
take the mass function to be of the form $m=\tilde{p}\sf{r}^{2}$
where $\tilde{p}$ is an arbitrary constant. For this example we
consider a particular form of the metric coefficients given in
$(\ref{abmet})$. We choose the coefficients to be separable in $r$
and $t$ so that
\begin{equation}
A = a(r), \quad \quad \quad \quad B = b(r)R(t)\label{metcoform1}
\end{equation}
The field equations $(\ref{0sheareinfield})$ then yield the
quantities
\begin{subequations}
\label{pertquants}
\begin{eqnarray}
\mu &=&
\frac{1}{R^2}\left[\frac{3}{a^2}\dot{R}^2-\frac{1}{b^2}\left(\frac{2b^{\prime
\prime}}{b}-\frac{b^{\prime
2}}{b^2}+\frac{4b^{\prime}}{rb}\right)\right]\label{pertrho}\\
p &=& \frac{1}{R^2}\left[-\frac{1}{a^2}\left(2R\ddot{R}+\dot{R}^2
\right)+\frac{1}{b^2}\left(\frac{b^{\prime
2}}{b^2}+\frac{2a^{\prime}b^{\prime}}{ab}+\frac{2}{r}\left(\frac{a^{\prime}}{a}+\frac{b^{\prime}}{b}
\right)\right)\right]\label{pertp}\\
q &=& -\frac{2a^{\prime}\dot{R}}{R^3 a^2 b^2}\label{pertq}
\end{eqnarray}
\end{subequations}
and the condition for the isotropy of pressure
\begin{equation}
\frac{a^{\prime \prime}}{a}+\frac{b^{\prime
\prime}}{b}-2\frac{b^{\prime 2}}{b^2}
-2\frac{a^{\prime}b^{\prime}}{ab}-\frac{a^{\prime}}{ra}-\frac{b^{\prime}}{rb}
= 0\label{pic}
\end{equation}
The boundary condition $(\ref{res2})$ now yields at $r=r_{\Sigma}$
the equation
\begin{equation}
2 R \ddot{R}+ {\dot{R}} ^2 + {\cal M}\dot{
R}={\cal N}+{\cal P}\label{expbndcon}
\end{equation}
where we have used the particular choice of the mass function $m=\tilde{p}\sf{r}^{2}$, the
potentials $(\ref{metcoform1})$ and the system $(\ref{pertquants})$.
In the above we have set
\begin{subequations}
\label{boundcoeffs}
\begin{eqnarray}
\cal M &=& -\frac{2a^{\prime}}{b}\label{boundcoeff1}\\
\cal N &=& \frac{a^2}{b^2}\left[\frac{b^{\prime
2}}{b^2}+2\frac{a^{\prime
}b^{\prime}}{ab}+\frac{2}{r}\left(\frac{a^{\prime}}{a}+\frac{b^{\prime}}{b}
\right)\right]\label{boundcoeff2}\\
\cal P &=& 6\frac{a^2 \tilde{p}}{b^2}\label{boundcoeff3}
\end{eqnarray}
\end{subequations}
which are constants in the integration process as
$(\ref{expbndcon})$ holds on the boundary $\Sigma$.

Equation $(\ref{expbndcon})$ does not have a general solution in
closed form; for the purpose of our investigation we observe that it
admits an elementary particular solution of the form
\begin{equation}
 R (t)=-{\cal C} t\label{specialsol}
\end{equation}
where $\cal C$ is given by
\begin{equation}
{\cal C}=\frac{1}{2}\left[{\cal M}\pm \left({\cal M}^2 +4\left({\cal
N}+{\cal P}\right)\right)^{\frac{1}{2}}\right]\nonumber
\end{equation}
This solution is useful in that at the surface of the collapsing
star the ratio $\frac{m_{\Sigma}}{{\sf{r}}_{\Sigma}}$ is independent
of time. The mass profile given by $(\ref{res1})$ and the solution
$(\ref{specialsol})$ yield the following expression
\begin{equation}
\frac{2m_{\Sigma}}{{\sf{r}}_{\Sigma}}=2\left[\frac{r^2b^2{\cal
C}^2}{2a^2}-\frac{rb^{\prime}}{b} -\frac{r^2b^{\prime
2}}{2b^2}\right]_{\Sigma}\label{schradius}
\end{equation}
which is valid on the stellar surface. It is clear that the
quantities in the above equation are evaluated at the boundary
$\Sigma$; there is sufficient freedom provided by $r$, $a$, $b$ and
$\cal C$ so that the right hand side can be constrained in such a
manner that the ratio $\frac{2m_{\Sigma}}{{\sf{r}}_{\Sigma}}$ is
strictly less than unity. Consequently no horizon will form. We can
illustrate this explicitly by setting $b(r)=1$ and
$a(r)=\frac{1}{2}\xi r^2 +\beta $ where $\beta$ and $\xi$ are
arbitrary constants. Then we get
\begin{equation}
1-\frac{2m_{\Sigma}}{{\sf{r}}_{\Sigma}}=\left[1-\frac{{\cal C}^2
r^2}{\left(\beta + \frac{1}{2}\xi r^2\right)^2}\right]_{\Sigma}
\end{equation}
We choose ${\cal C}^2$ to be less than $\frac{\beta ^2}{r^2}+\beta
\xi + \frac{1}{4}\xi ^2 r^2$ on $\Sigma$ so that
$1-\frac{2m_{\Sigma}}{{\sf{r}}_{\Sigma}}$ is not zero for all time.
Consequently the boundary surface never reaches the horizon. This
simple example demonstrates the absence of the horizon and is
similar to the result of Banerjee et al. \cite{b38} which holds in
the conventional Vaidya spacetime. We regain their results when
${\cal P}=0$ in $(\ref{boundcoeffs})$. We interpret our result to
mean that there is no accumulation of energy in the interior of the
radiating star as it releases energy at the rate of generation
during the collapse.

\section{Conclusion}
In this work we have produced a general model of a relativistic
radiating star by performing the smooth matching of a shear-free
interior spacetime to the generalised Vaidya exterior spacetime,
across a timelike spatial hypersurface. We have demonstrated that
with the generalised Vaidya radiating metric, the junction
conditions on the stellar surface change substantially, and
consequently represents a more general atmosphere surrounding the
star. The atmosphere is a superposition of null dust and a string
fluid. We find that the density of the string fluid affects the
pressure at the stellar boundary. We have shown explicitly that
\begin{equation}
p=qB-\rho_{string}\nonumber
\end{equation}
at the stellar surface. If the weak and strong energy conditions or
the dominant energy conditions are satisfied then $\rho_{string}\geq
0\quad(\mu\neq0)$ and $\rho_{string}\geq P_{string}\geq
0\quad(\mu\neq0)$ respectively. This indicates that for outgoing
heat flux in gravitational collapse, the exterior string fluid
density reduces the pressure on the stellar boundary. It is
interesting to note that we have shown using a geometric approach
that the derivative of the mass function with respect to the
exterior radial coordinate is related to the string density.
By means of a simple example we have considered the formation of horizons
for a heat conducting sphere which radiates energy into the
generalised Vaidya external atmosphere during collapse. The horizon
does not appear in this example at any stage of the collapse.

\begin{acknowledgements}
MG and GG thank the National Research
Foundation and the University of KwaZulu-Natal for financial
support. SDM acknowledges that this work is based upon research
supported by the South African Research Chair Initiative of the
Department of Science and Technology and the National Research
Foundation. The authors would also like to thank Sanjay Wagh and
Subharthi Ray for constructive criticisms and useful discussions.
\end{acknowledgements}


\begin{thebibliography}{}


\bibitem{b1}
Di Prisco, A., Herrera, L., Le Denmat, G.,  MacCallum, M.A.H., Santos, N.O.:
Phys. Rev. D \textbf{76}, 064017 (2007)

\bibitem{b2}
Govender, M., Maharaj, S. D., Maartens, R.: Class. Quantum
Grav. \textbf{15}, 323 (1998)

\bibitem{b3}
Herrera, L.,Le Denmat, G., Santos, N.O.: Phys. Rev. D
\textbf{79}, 087505 (2009)

\bibitem{b4}
Pinheiro, G., Chan, R.: Gen. Relativ. Gravit. \textbf{40},
2149 (2008)

\bibitem{b5}
Goswami, R., Joshi, P.S.: Phys. Rev. D \textbf{76}, 084026
(2007)

\bibitem{b6}
Joshi, P.S.: Pramana - J. Phys. \textbf{69}, 119 (2007)

\bibitem{b7}
Madhav, T.A.,  Goswami, R., Joshi, J.S.: Phys. Rev. D
\textbf{72}, 084029 (2005)

\bibitem{b8}
Santos, N.O.: Mon. Not. R. Astron. Soc. \textbf{216}, 403
(1985)

\bibitem{b9}
Kramer, D.: J. Math. Phys. \textbf{33}, 1458 (1992)

\bibitem{b10}
Maharaj, S. D., Govender, M.: Aust. J. Phys. \textbf{50},
959 (1997)

\bibitem{b11}
Kolassis, C.A., Santos, N.O., Tsoubelis, D.: Astrophys. J.
\textbf{327}, 755 (1988)

\bibitem{b12}
Thirukkanesh, S., Maharaj, S.D.: J. Math. Phys.
\textbf{50}, 022502 (2009)

\bibitem{b13}
de Oliviera, A.K.G., Santos, N.O, Kolassis, C.A.: Mon.
Not. R. Astron. Soc. \textbf{216}, 1001 (1985)

\bibitem{b14}
Nogueira, P.C., Chan, R.: Int. J. Mod. Phys. D \textbf{13},
1727 (2004)

\bibitem{b15}
Herrera, L., Le Denmat, G., Santos, N.O., Wang, A.: Int. J.
Mod. Phys. D \textbf{13}, 583 (2004)

\bibitem{b16}
Herrera, L.,  Di Prisco, A., Ospino, J.: Phys. Rev. D
\textbf{74}, 044001 (2006)

\bibitem{b17} Maharaj, S.D.,  Govender, M.: Int. J. Mod. Phys.
D \textbf{14}, 667 (2005)

\bibitem{b18}
 Misthry, S.S.,  Maharaj, S.D.,  Leach, P.G.L.: Math. Meth.
Appl. Sci. \textbf{31}, 363 (2008)

\bibitem{b19}
Naidu, N.F., Govender, M.,  Govinder, K.S.: Int. J. Mod.
Phys. D \textbf{15}, 1053 (2006)

\bibitem{b20}
Rajah, S.S., Maharaj, S.D.: J. Math. Phys. \textbf{49},
012501 (2008)

\bibitem{b21}
Pinheiro, G.,  Chan, R.: Gen. Relativ. Gravit. \textbf{43},
1451 (2011)

\bibitem{b22}
 Husain, V.: Phys. Rev. D \textbf{53}, 1759 (1996)

\bibitem{b23}
 Wang, A., Wu,Y.: Gen. Relativ. Gravit. \textbf{31}, 107
(1999)

\bibitem{b24}
 Glass, E.N.,  Krisch, J.P.: Phys. Rev. D \textbf{57}, 5945
(1998)

\bibitem{b25}
 Glass, E.N., Krisch, J.P.: Class. Quantum Grav.
\textbf{16}, 1175 (1999)

\bibitem{b26}
Krisch, J.P., Glass, E.N.: J. Math Phys. \textbf{46},
062501 (2005)

\bibitem{b27}
Lichnerowicz, A.: Theories Relativistes de la Gravitation
et de l'Electromagnetisme. Masson, Paris (1955)


\bibitem{b28}
O' Brien, S.,  Synge, J.L.: Dublin Inst. Adv. Stud. A
\textbf{9}, 1 (1952)

\bibitem{b29}
 Lake, K., Zannias, T.: Phys. Rev. D \textbf{43}, 1798
(1991)

\bibitem{b30}
 Poisson, E., Israel, W.: Phys. Rev. D \textbf{41}, 1796
(1990)

\bibitem{b31}
 Hawking, S.W., Ellis, G.F.R.: The large scale structure
of spacetime. Cambridge University Press, Cambridge (1973)

\bibitem{b32}
 Barriola, M., Vilenkin, A.: Phys. Rev. Lett. \textbf{63},
341 (1989)

\bibitem{b33}
 Chamorro, A.,  Virbhadra, K.S.: Pramana - J. Phys.
\textbf{45}, 181 (1995)

\bibitem{b34}
 Virbhadra, K.S.:  Phys. Rev. D \textbf{41}, 1086 (1990)

\bibitem{b35}
Virbhadra, K.S.: Phys. Rev. D \textbf{42}, 2919 (1990)

\bibitem{b36}
Virbhadra, K.S.: Phys. Rev. D \textbf{60}, 104041 (1999)

\bibitem{b37}
 Yang, I.C.: Chin. J. Phys. \textbf{45}, 497 (2007)

\bibitem{b38}
 Banerjee, A., Chatterjee, S., Dadhich, N.K.: Mod. Phys. Lett. A
\textbf{17}, 2335 (2002).











\end{thebibliography}
\end{document}